\documentstyle[multicol,aps,prl,twocolumn,epsfig]{revtex}
\newcommand{\bm}[1]{\mbox{\boldmath $#1$}}
\begin{document}
\draft %%%%%
\twocolumn[\hsize\textwidth\columnwidth\hsize\csname@twocolumnfalse\endcsname

\title{Universal decay of scalar turbulence} \author{M.~Chaves$^{1}$,
G.~Eyink$^{2}$, U.~Frisch$^{1}$ and M.~Vergassola$^{1}$\\ \small{$^{1}$ CNRS,
Observatoire de la C\^ote d'Azur, B.P. 4229, 06304 Nice Cedex 4,
France.}\\ \small$^2$ Department of Mathematics, Univ. of Arizona,
Tucson, AZ85721, USA.}
%\begin{document}
\date{\today}
\maketitle
\begin{abstract}
The asymptotic decay of passive scalar fields is solved analytically
for the Kraichnan model, where the velocity has a short correlation
time. At long times, two universality classes are found, both
characterized by a distribution of the scalar -- generally
non-Gaussian -- with global self-similar evolution in time. Analogous
behavior is found numerically with a more realistic flow resulting
from an inverse energy cascade.
\end{abstract}
\pacs{PACS number(s)\,: 47.10.+g, 47.27.-i, 05.40.+j}]

Recent research on passive scalars in prescribed turbulent flows has
mostly concentrated on universal small-scale properties in the
presence of large-scale random pumping of the scalar (see
Ref.~\cite{SS2000} and references therein). Another important problem
is the decay of a scalar field $\theta({\bm r},t)$ with prescribed
random initial conditions. It is then possible to obtain a stronger
form of universality {\em at all scales}, but only at long times, when
(most) of the information in the initial conditions has been washed
out. Specifically, we shall be interested in self-similar decay at
infinite (or very large) P\'eclet numbers.  We ask the following
questions: (i) Is it possible to find time-dependent rescaling factors
of length and scalar amplitude such that in the rescaled variables the
full single-time probability of the passive field becomes independent
of time as $t\to \infty$ ? (ii) If the first question holds true, what
are the universality classes (with respect to the initial conditions)
of the asymptotically self-similar scalar field and what are their
statistical properties? Such issues are of course not limited to
passive scalars. For Burgers turbulence, a systematic theory of
self-similar decay giving, e.g., the law of decay of the energy can be
derived \cite{Kida79,Gurbetal97,FM2000}.  We shall return to the
corresponding issues for 3-D Navier--Stokes turbulence in the
conclusions.  Here we shall first address such issues within the
framework of the Kraichnan model\cite{rhk94} of passive scalars in
which the velocity has a very short correlation time. This allows
analytical and numerical work shedding light on the basic mechanisms
of scalar decay. We show that there are two universality classes, both
generally having self-similar non-Gaussian distributions. Spatially
smooth and very rough velocity fields represent two limiting special
cases: for the former self-similar decay is broken \cite{Son98,BF99}
while for the latter nearly Gaussian self-similar decay obtains. We
then turn, using numerical simulations, to decay in the presence of
more realistic velocity fields for a scalar transported by a 2-D
velocity field arising from an inverse energy cascade process
\cite{rhk67}. This also leads to self-similar non-Gaussian decay with
exponential tails in the single-point scalar pdf.

The advection-diffusion equation governing the evolution of a passive
scalar field is
\begin{equation}
\label{passive}
\partial_t\theta({\bm r},t)+{\bm v}({\bm r},t)\cdot{\bm \nabla}\,
\theta({\bm r},t)=\kappa\nabla^2\theta({\bm r},t) ,
\end{equation}
where ${\bm v}({\bm r},t)$ is the incompressible advecting flow and
$\kappa$ is the molecular diffusivity.  Eq.~(\ref{passive})
governs the evolution of the probability density at position ${\bm r}$
and time $t$ of tracer particles. Their Lagrangian trajectories ${\bm
\rho}(t)$ obey the stochastic equation
\begin{equation}
d{\bm \rho}(t)={\bm v}({\bm \rho}(t),t)\,dt+\sqrt{2\kappa}\,\,d{\bm
\beta}(t),
\label{Lagr}
\end{equation}
where ${\bm \beta}(t)$ is the $d$-dimensional isotropic Brownian
motion.  The scalar correlation functions at time $t$ can then be 
expressed in terms of the initial ones as
\begin{eqnarray}
\label{correlation}
C_N(\underline{\bm r},t)&=&\langle\theta({\bm r}_1,t)\ldots
\theta({\bm r}_N,t)\rangle \nonumber \\ &=& \int {\cal
P}_N\left(\underline{\bm \rho},0|\underline{\bm
r},t\right)\,C_N(\underline{\bm \rho},0)\,d\underline{\bm \rho}.
\end{eqnarray}
Here, $\underline{\bm r}$ denotes the set ${\bm r}_1,\ldots,{\bm r}_N$
and ${\cal P}_N$ is the probability density that $N$ Lagrangian
particles, being at $\underline{\bm r}$ at time $t$, were previously
at $\underline{\bm \rho}$ at time $t=0$. The initial condition
$\theta({\bm r},0)$ is taken random and statistically homogeneous with
correlation length $L$. The statistical homogeneity implies that the
absolute position of the particles is irrelevant and only the
separation variables ${\bm r}_{i,j}={\bm r}_i-{\bm r}_j$ matter. In
the sequel we shall always consider the propagator in the restricted
homogeneous sector, i.e. in the $(N-1)d$ separation variables.  Note
that the evolution equation (\ref{passive}) conserves the so-called
Corrsin integral \cite{Corr51}
\begin{equation}
\label{Corrsin}
J_0=\int C_2({\bm r},t)\,d{\bm r}.
\end{equation}
That conservation law alone does not however allow to predict the
scalar energy decay:  for this, we also need to know the
growth of the scalar correlation length with time. This is where the
specific properties of the advecting velocity come into play.

Let us begin with the Kraichnan model: the
velocity is statistically homogeneous and stationary, isotropic and
Gaussian. The second-order correlation function of its increments is
\begin{eqnarray}
\label{correlations}
D_{\alpha\beta}({\bm r},t)\equiv \langle v_{\alpha}({\bm
r},t)\,v_{\beta}({\bm r},0) \rangle -\langle v_{\alpha}({\bm
r},t)\,v_{\beta}({\bm 0},0) \rangle = \nonumber \\
D r^{-\gamma}\left[r^2\left(\gamma+d+1\right)\delta_{\alpha\beta}-(2-\gamma)\,
r_{\alpha}r_{\beta}\right]\delta(t)\, ,
\end{eqnarray}
where $\gamma$ controls the roughness of the field.  Smooth flows
correspond to $\gamma=0$, while for $\gamma=2$ the opposite limit of
very rough fields is achieved.  Hereafter, we shall concentrate on the
generic non-smooth case $\gamma\neq 0$. The growth of the scalar
length-scale is determined dimensionally to be $L(t)\sim t^{1/\gamma}$
and the units are chosen for convenience to set $L=D=1$.  Note that
the velocity field in the Kraichnan model is statistically invariant
under time inversion, i.e. ${\bm v}({\bm r},t)$ has the same
statistical properties as $-{\bm v}({\bm r},-t)$. It follows that the
$N$-particle propagator satisfies
\begin{equation}
\label{inversion}
{\cal P}_N\left(\underline{\bm \rho},0|\underline{\bm
r},t\right)={\cal P}_N\left(\underline{\bm \rho},t|\underline{\bm
r},0\right)={\cal P}_N\left(\underline{\bm r},t|\underline{\bm
\rho},0\right) .
\end{equation}
The major consequence of the short correlation time assumption is that
the Lagrangian trajectories are a Markov stochastic process and that
the $N$-particle propagators ${\cal P}_N$ obey Fokker-Planck
equations (see, e.g., Ref.~\cite{SS2000}):
\begin{equation}
\label{Pn}
\partial_t {\cal P}_N+\sum_{i\neq
j}\nabla^{r_i}_{\alpha}\nabla^{r_j}_{\beta}\left[\kappa\delta_{\alpha\beta}+
{1\over 2}D_{\alpha\beta}({\bm r}_{i,j})\right]{\cal P}_N=0.
\end{equation}
The advective term in (\ref{passive}) conserves all the moments
$\langle\theta^n\rangle$.  The scalar decay is therefore due to the
effects of the molecular dissipation and, as usual in turbulence, we
are interested in the singular limit $\kappa\to 0$ (infinite P\'eclet
numbers). The scalar dissipation
$\kappa\langle\left(\nabla\theta\right)^2\rangle$ remains indeed
finite in the limit. From the Lagrangian point of view, the molecular
noise in (\ref{Lagr}) can be neglected and the singularity of the
limit shows up in the fact that initially coinciding particles
separate in finite time \cite{BGK,FMV}.  For $\kappa\to 0$, the
fundamental solution of (\ref{Pn}) is self-similar and the propagator
possesses the scaling property
\begin{equation}
{\cal P}_N(\lambda \underline{\bm \rho},0|\lambda\underline{\bm r},
\lambda^{\gamma}t) \ =\ \lambda^{-(N-1)d}\,\,{\cal P}_N(\underline{\bm
\rho},0|\underline{\bm r},t)\,.
\label{Nscal}
\end{equation}

The simplest objects to investigate are the single-point moments
$\langle \theta^{2n}\rangle(t)$ and we are interested in their
long-time behavior $t\gg 1$.  The first universality class is formed
by the initial conditions where the Corrsin invariant $J_0$ defined in
(\ref{Corrsin}) does not vanish. We shall also make the natural
assumption that the initial scalar correlations decay rapidly enough
to ensure integrability of the cumulants of arbitrary order.  The
long-time behavior of the single-point moments is obtained by setting
${\bm r}=0$ in (\ref{correlation}) and using (\ref{Nscal}) to rescale
out the time dependence.  The resulting object involves
$t^{d/\gamma}C_2({\bm \rho} t^{1/\gamma},0)$ which tends to
$\delta({\bm \rho})J_0$.  The final result is
\begin{eqnarray}
\label{blues2}
\langle \theta^{2n}\rangle(t)&\simeq&{J_0^n\over t^{nd/\gamma}} (2n-1)!!
\nonumber \\ &\times & \int {\cal P}_{2n}\left({\bm \rho}_1,{\bm
\rho}_1,\ldots {\bm \rho}_n,{\bm \rho}_n,1|\underline{\bm
0},0\right)\,d{\bm \rho}_1\ldots d{\bm \rho}_n ,
\end{eqnarray}
as is easily seen for the special case of initial Gaussian statistics.
In fact, it is not hard to show that the same 
%%%%%%%%%%%%%%%%%%%%%%%%%%%%%%%%%%%%%%%%%%%%%%%%%%%
%%%%%%%%%%%%
\narrowtext
\begin{figure}
\epsfxsize=7.6truecm
\epsfbox{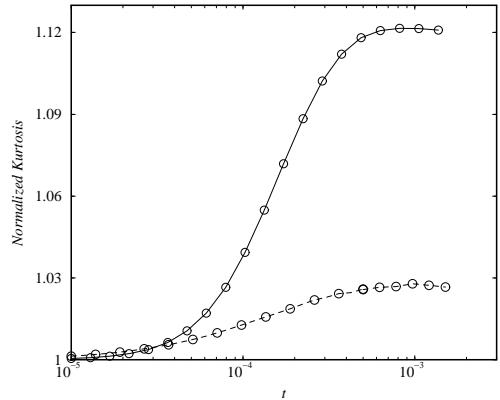}
\caption{The normalized kurtosis $\langle\theta^4\rangle/
3\langle\theta^2\rangle^2$ {\it vs} time in the
Kraichnan model for the two values $\gamma=5/4$ (dashed line) and
$\gamma=2/3$ (solid line). The error bars are comparable to the 
size of the symbols.}
\label{f1}
\end{figure}
%%%%%%%%%%%%%%%%%%%%%%%%%%%%%%%%%%%%%%%%%%%%%%%%%%%
%%%%%%%%%%%%%
\noindent result holds for all initial data in the universality class
$J_0\neq 0$ and rapid correlation decay. The integrability condition
on the cumulant of order $2n$ implies indeed that it must depend on
all the $(2n-1)$ independent interparticle separations. In the limit
of long times, each dependence brings a function $\delta({\bm
\rho}_{ij})$ and a factor $t^{d/\gamma}$ in the denominator.  The total
non-Gaussian contribution from the initial conditions will therefore
decay faster than $t^{-nd/\gamma}$. Note that this argument implies a
stronger form of universality than in forced steady states: the
asymptotic pdf is here independent of the initial data, whereas in the
steady states the scalar pdf depends on the details of the force.  The
expression (\ref{blues2}) shows that the scalar statistics is
self-similar in time. For example, the single point pdf ${\cal
P}(\theta,t)$ takes the form $t^{d/2\gamma}Q(\theta t^{d/2\gamma})$.
The propagator in the limit of initially coinciding points that
appears in (\ref{blues2}) is a regular function of its arguments
\cite{BGK}. The smooth case $\gamma=0$ is special in this respect
since nearby particles separate exponentially and their separation is
proportional to its initial value. This is the physical reason why
(\ref{blues2}) and the ensuing self-similarity of the scalar
distribution do not hold for $\gamma=0$ (see Refs.~\cite{Son98,BF99}).
Note that the asymptotic pdf ${\cal P}(\theta,t)$, although universal
and self-similar, is generally non-Gaussian. From (\ref{blues2}), the
hypothesis of Gaussianity is equivalent to the factorization of the
probability for the $2n$ particles to collapse into pairs, i.e. the
integral in (\ref{blues2}) should be equal to $\left[\int {\cal
P}_2\left({\bm \rho},{\bm \rho},1|\underline{\bm 0},0\right)\,d{\bm
\rho}\right]^n$.  This is generally not the case, due to the
correlations existing among the particle trajectories. The only
exception is $\gamma=2$ where the particles are independent. The
degree of non-Gaussianity is thus expected to decrease when $\gamma$
increases, as confirmed in Fig.~1.  The simulations were performed by
the Lagrangian method presented in Refs.~\cite{FMV,FMNV}.

The second universality class is for initial conditions with
Corrsin invariant $J_0=0$.  From the realizability of 
%%%%%%%%%%%%%%%%%%%%%%%%%%%%%%%%%%%%%%%%%%%%%%%%%%%
%%%%%%%%%%%%
\narrowtext
\begin{figure}
\epsfxsize=7.6truecm
\epsfbox{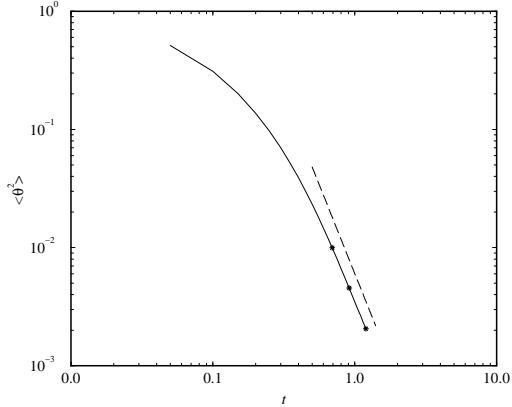}
\caption{The scalar energy $\langle\theta^2\rangle$ {\it vs} $t$ for
the advection by a 2D flow generated by an inverse energy cascade.
The dashed line is the asymptotic power law behavior $1/t^3$.}
\label{f2}
\end{figure}
%%%%%%%%%%%%%%%%%%%%%%%%%%%%%%%%%%%%%%%%%%%%%%%%%%%
%%%%%%%%%%%%%
\noindent the field it follows that the integral $J_1=\int
r^{\gamma}C_2({\bm r})\,d{\bm r}$ is non-vanishing \cite{Greg}. It is
also easily checked using (\ref{correlation}) and (\ref{Pn}) that the
integral remains invariant in time. $J_0=0$ implies that the integral
of the correlation $C_N$ with respect to any of its arguments
vanishes.  Furthermore, the single-point moments can then be expressed
as
\begin{equation}
\label{J1}
\langle\theta^{2n}\rangle (t)=\int {\cal Q}_{2n}\left(\underline{\bm
\rho},t|\underline{\bm r},0\right)\,C_{2n}\left(\underline{\bm
\rho},0\right)\,d\underline{\bm \rho}.
\end{equation}
\noindent Here
\begin{equation}
{\cal Q}_{2n}\left(\underline{\bm \rho},t|\underline{\bm r},0\right)
\equiv \left(\prod_{i=2}^{2n}\Delta_{{\bm \rho}_{i,1},{\bm
0}}\right){\cal S}_{2n} {\cal P}_{2n}\left(\underline{\bm
\rho},t|\underline{\bm r},0\right),
\label{differenza}
\end{equation}
${\cal S}_{2n}$ denotes the symmetrizer in the $\underline{\bm \rho}$
variables and the difference operator $\Delta$ acts on a function
$f({\bm x})$ as $\Delta_{{\bm x},{\bm y}}f({\bm x})=f({\bm x})-f({\bm
y})$. To obtain (\ref{J1}), we set $\underline{\bm r}=0$ in
(\ref{correlation}), symmetrize in the $\underline{\bm \rho}$
variables and use the condition $J_0=0$ to reexpress the result in
terms of difference operators. For this, note that whenever any of the
${\bm \rho}$ variables in the propagator is replaced by ${\bm 0}$ the
integral vanishes.  The long-time behavior of the single-point moments
is then derived from (\ref{J1}) exploiting again the scaling property
(\ref{Nscal}).  For Gaussian initial conditions, we obtain
\begin{eqnarray}
\label{blues3}
&&\langle \theta^{2n}\rangle (t)\simeq{J_1^n\over t^{n(d/\gamma+1)}}
(2n-1)!!  \times \\ && \lim \int {{\cal Q}_{2n}\left({\bm \rho}_1,{\bm
\rho}_2,\ldots {\bm \rho}_{2n-1},{\bm \rho}_{2n},1|\underline{\bm
0},0\right)\over \left[\rho_{1,2}\ldots
\rho_{2n-1,2n}\right]^{\gamma}}\,\prod_{i=1}^n d{\bm \rho}_{2i-1}\nonumber ,
\end{eqnarray}
where we used the convergence of $\lambda^{d/\gamma}\left(\lambda
\rho\right)^{\gamma}C_2(\lambda {\bm \rho},0)$ to $\delta({\bm
\rho})J_1$ for large $\lambda$'s. The limit of
$\rho_{1,2}\ldots\rho_{2n-1,2n}$ tending to zero, which is understood
in (\ref{blues3}), has been shown to be well defined \cite{Anti}. The
strong universality of the scalar distribution is proved as for the
$J_0\neq 0$ case. The expression (\ref{blues3}) indicates that the
asymptotic scalar distribution is again self-similar although the
rescaling factor is now $t^{d/(\gamma+1)}$, in agreement with the
exact solution for the second-order case \cite{Greg}.

The long-time behavior of multi-point scalar correlations can be
derived from (\ref{correlation}) as for the single-point
moments. Using the scaling property (\ref{Nscal}) we are generally
led to analyze ${\cal P}_N(\underline{\bm \rho},1|\underline{\bm
r}/t^{1/\gamma},0)$, i.e.  the limit where the $N$ particles are
initially close. This limit is crucial also for the small-scale
scaling properties in the presence of pumping and leads to an asymptotic 
expansion for the propagator~\cite{BGK}
\begin{equation}
{\cal P}_{_N}(\underline{\bm \rho},t|\lambda\underline{\bm r},0) \simeq
\sum\limits_{{a,k}}\,\lambda^{p_{a} +\gamma k}\
f_{{a},k}(\underline{\bm r})\ \,g_{{a},k}(\underline{ \bm \rho},t),
\label{asex}
\end{equation}
valid for small $\lambda$. \,The first sum is over the zero modes
$\,f_{{a}}\equiv f_{{a},0}\,$ with scaling dimensions $\,p_{a}$ (in
increasing order).  The functions $\,f_{{a},k}\,$ for different
$\,k\,$ form the corresponding towers of the so-called slow modes.
The leading term in the expansion (\ref{asex}) comes from the constant
zero mode $\,f_{0,0}=1\,$ which corresponds to $\,g_{0,0}(\underline{
\bm \rho},t)={\cal P}_{_N}(\underline{\bm \rho},t |\underline{\bm
0},0)$, i.e. the propagator for $N$ particles that are initially at
the same position.  Inserting (\ref{asex}) into (\ref{correlation}) and
using the dominance of $g_{0,0}$ is is easy to check that generic
multi-point correlations will asymptotically decay as the
corresponding single-point moment. Sub-dominant terms emerge by
combining the correlation functions to cancel the dominant
contribution.  The classical example is given by the structure
functions $S_{2n}({\bm r},t)=\langle [\theta({\bm r},t)-\theta({\bm
0},t)]^{2n}\rangle$, related to the correlation functions by
\begin{equation}
S_{2n}({\bm r},t)=\left(\prod_{i=1}^{2n}\Delta_{{\bm r}_i,{\bm
0}}\right) C_{2n}(\underline{\bm r},t)\,.
\label{soleil}
\end{equation}
The difference operator $\Delta$ was defined in
(\ref{differenza}). The crucial point to remark is that all the
reducible contributions to the correlations, i.e. not depending on all
the coordinates ${\bm r}_1,\ldots {\bm r}_{2n}$, are annihilated by
the combination of the structure functions. Inserting (\ref{asex})
into (\ref{correlation}) we obtain
\begin{eqnarray}
S_{2n}&\simeq&{\left(\prod_{i=1}^{2n}\Delta_{{\bm r}_i,{\bm
0}}\right) f_{2n,0}(\underline{\bm r})\over t^{p_{2n}\over\gamma}}
\int g_{2n,0}(\underline{\bm
\rho},1)\,\,C_{2n}(t^{1\over\gamma}\underline{\bm
\rho},0)\,\,d\underline{\bm \rho} \nonumber \\ &\propto &
\left({r\over t^{1/\gamma}}\right)^{\zeta_{2n}}
\langle\theta^{2n}\rangle(t)\,,
\end{eqnarray}
where $f_{2n,0}$ is the irreducible zero mode in (\ref{asex}) with the
lowest dimension. The quantities $\zeta_{2n}=p_{2n}$ coincide with the
exponents of the structure functions in forced steady states
\cite{BGK}. This justifies the classical notion of quasi-equilibrium,
i.e. that short-distance scaling laws are the same in forced and
decaying situations provided the basic parameters are made
appropriately time-dependent.  

The analytical solution found for the Kraichnan model permits to
capture the basic features of the scalar decay. It is in particular
quite clear that the key ingredient for the asymptotic self-similarity
of the scalar pdf is the validity of the rescaling property
(\ref{Nscal}). We also performed
%%%%%%%%%%%%%%%%%%%%%%%%%%%%%%%%%%%%%%%%%%%%%%%%%%%
%%%%%%%%%%%%
\narrowtext
\begin{figure}
\epsfxsize=8.0truecm
\epsfbox{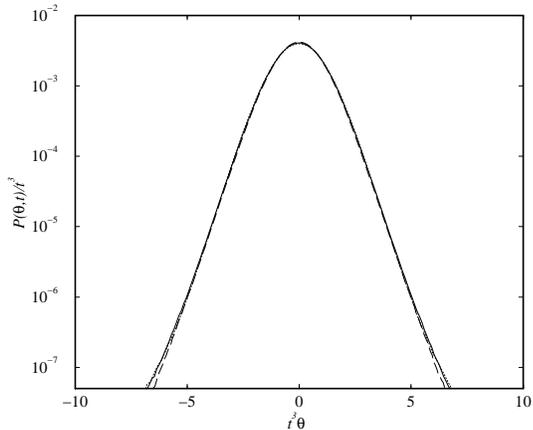}
\caption{The rescaled scalar pdf's {\it vs} $t^3\theta$ for the
inverse energy cascade flow at the three times marked by a star in
Fig.~2.  The values on the abscissa are normalized by their
r.m.s. values.}
\label{f3}
\end{figure}
%%%%%%%%%%%%%%%%%%%%%%%%%%%%%%%%%%%%%%%%%%%%%%%%%%%
%%%%%%%%%%%%%
\noindent   simulations using a more realistic
``inverse energy cascade'' 2-D velocity field which show that such property
is not limited to short-correlated velocities and is expected to hold
(at least for large enough times) for any self-similar velocity
field. Indeed the ``inverse cascade'' flow is scale-invariant (no
intermittency) of exponent $1/3$, isotropic and, most important, has
finite realistic correlation times \cite{SY,PT98,BCV99}.  The scalar
initial distribution is taken Gaussian with correlations decaying
exponentially and a correlation length comparable to the velocity
injection scale (see Ref.~\cite{BCV99} for more details). In Fig.~2 we
present the curve of the energy decay showing that the exponent
coincides with the value $d/\gamma$ obtained from the Kraichnan model
($\gamma=2/3$ for the inverse cascade flow). The rescaled scalar pdf's
${\cal P}(\theta,t)/t^3$ for three times in the asymptotic power-law
region are presented in Fig.~3. It is quite clear that the asymptotic
distribution is again self-similar and its tails are exponential.  The
tails are indeed dominated by the rare events governed by Poisson
statistics where the scalar is weakly mixed by diffusive effects
\cite{SS94}.

For passive scalar decay in an intermittent velocity field, such as a
3-D turbulent flow, the presence of various scaling exponents makes it
unlikely that the propagator possesses a rescaling property like
(\ref{Nscal}).  The self-similarity of the scalar decay found here
might then be destroyed, with the kurtosis and higher-order scalar
moments possibly growing indefinitely with time.  An even more
challenging problem is that of the decay of 3-D turbulent flow which
has been frequently addressed, but usually in a phenomenological way
and with the emphasis on the law of energy decay (see
Refs.~\cite{MY,UF95} for a review). Is there self-similar and
universal decay as for the problem considered in this paper or is the
kurtosis of the velocity growing indefinitely, possibly leading to a
highly intermittent spatial distribution of the kinetic energy? It is
of great interest to address such issues experimentally and by
advanced direct numerical simulations.

\medskip \noindent 
We are grateful to J.~Bec, A.~Celani, M.~Cher\-tkov,
G.~Falkovich, K.~Gaw\c{e}dzki, B.~Shraiman and K.R.~Sreenivasan for
useful discussions. The support by the European Union under Contract
HPRN-CT-2000-00162 and by the National Science Foundation under Grant
No. PHY94-07194 is acknowledged. Simulations were performed at IDRIS
(no~991226).

\end{document}